\documentclass[aps,prb,floatfix,twocolumn,superscriptaddress,showpacs,amsmath,amssymb]{revtex4}
\usepackage{graphicx}
\usepackage{epstopdf}
\usepackage{epsfig} 
\allowdisplaybreaks

\newcommand{\be}{\begin{equation}}
\newcommand{\ee}{\end{equation}}
\newcommand{\bea}{\begin{eqnarray}}
\newcommand{\eea}{\end{eqnarray}}
\newcommand{\ba}{\begin{eqnarray*}}
\newcommand{\ea}{\end{eqnarray*}}
\newcommand{\dagga}{{\phantom{\dagger}}}

\newcommand{\bk}{\mathbf{k}}

\newcommand{\br}{\mathbf{r}}

\newcommand{\dis}{\displaystyle}

\newcommand{\up}{\uparrow}
\newcommand{\down}{\downarrow}
\newcommand{\fract}[2]{\frac{\dis #1}{\dis #2}}

\newcommand{\eqn}[1]{(\ref{#1})}

\begin{document}

\title{Lattice and surface effects in the out-of-equilibrium dynamics of the Hubbard model}

\author{Patrice Andr\'e} 
\affiliation{International
  School for Advanced Studies (SISSA), and CNR-IOM Democritos, 
  Via Bonomea 265, I-34136 Trieste, Italy} 
\author{M. Schir\'o}
\affiliation{Princeton
 Center for Theoretical Science and Department of Physics, Joseph Henry
  Laboratories, Princeton University, Princeton, NJ
  08544, USA}
\author{Michele Fabrizio} 
\affiliation{International School for
  Advanced Studies (SISSA), and CNR-IOM Democritos, Via Bonomea
  265, I-34136 Trieste, Italy} 
\affiliation{The Abdus Salam
  International Centre for Theoretical Physics (ICTP), P.O.Box 586,
  I-34014 Trieste, Italy} 


\date{\today} 


\begin{abstract}
We study, by means of the time-dependent Gutzwiller approximation, the out of equilibrium dynamics of a half-filled Hubbard-Holstein model of correlated electrons interacting with local phonons. Inspired by pump-probe experiments, where intense light pulses selectively induce optical excitations that trigger a transient out-of-equilibrium dynamics, here we inject energy in the Hubbard bands by a non-equilibrium population of empty and doubly-occupied sites. We first consider the case of a global perturbation, acting over the whole sample, and find evidence of a mean-field dynamical transition where the lattice gets strongly distorted above a certain energy threshold, despite the weak strength of the electron-phonon coupling by comparison with the Hubbard repulsion.\\
Next, we address a slab geometry for a correlated heterostructure and study the relaxation dynamics across the system when the perturbation acts locally on the first layer. While for weak deviations from equilibrium the excited surface is able to relax by transferring its excess energy to the bulk, for large deviations the excess energy stays instead concentrated into the surface layer. This 
self-trapping occurs both in the absence as well as in the presence of electron-phonon coupling. Phonons actually enforce the trapping by distorting at the surface.
\end{abstract}
\maketitle

\section{Introduction}

The transient dynamical behavior of correlated materials optically excited far from equilibrium is currently attracting growing interest, due to the impressive advances in time-resolved spectroscopy with femtosecond resolution. 
By shining the sample with intense ultra-fast pulses (pump), one can trigger non-equilibrium transient states, whose physical properties are then recorded by a second pulse arriving at fixed time delay (probe). The unique feature of these experimental techniques is to give access to dynamical information unavailable to conventional time-averaged frequency domain spectroscopies.\cite{Giannetti_NatComm11} 
In addition, when irradiation is sufficiently strong, one can even stabilize transient states with fundamentally different physical properties,\cite{Ichikawa_NatMat11} thus paving the way to a complete control of material properties by light.\cite{Fausti_Science11} As correlated electron systems are often on the verge of a Mott metal-to-insulator transition, this portends the utmost important possibility of optically manipulating their conducting properties on ultra-short time-scales.\cite{Dean_PRL11}

Motivated by these achievements, the research activity on transient ultrafast dynamics 
in correlated electronic systems has rapidly 
grown in recent years.\cite{Kollar_prb08,Kollar_prb09,Freericks_prl09,Moritz_prb10,Eckstein_prb11} From a theoretical perspective, one can expect a non trivial and rich transient dynamical behavior to emerge, reflecting the complex interplay between electrons, phonons, spins and orbital degrees of freedom that characterizes the phase diagram of these materials.
On a more fundamental level, the crucial question concerns whether these experiments could allow to explore novel metastable phases of correlated quantum matter that can only be accessed along non-thermal pathways.

A common wisdom is that the effect of perturbing the system 
by an ultra-short laser pulse can be qualitatively accounted for by an effective-temperature description.\cite{Parker_Tstar_PRB75,2Temp_Allen_PRL87,Perfetti_Georges_PRL06,T_eff_Tao_PRB10} Within this picture, the injected energy would turn first, on few femtoseconds, into heat for the electron sub-system only. At later times, picoseconds, the electronic heat is gradually transferred to the lattice so that, eventually, the whole system flows to a thermal state at higher temperature than the initial one. Under such a thermodynamic assumption, optical pumping should mimic the role of heating, hence allow accessing, possibly much faster, all phases that are reached upon raising temperature at equilibrium. In the specific case of correlated materials, this entails the possibility of photo-inducing 
metal-to-insulator transitions. Indeed, there exist many examples of 
Mott insulators that can be driven metallic upon increasing 
temperature, like e.g. V$_2$O$_3$\cite{V2O3,All-oxides} and VO$_2$\cite{All-oxides}, and, vice versa, metals that turn Mott insulating upon heating, like the same V$_2$O$_3$\cite{V2O3,All-oxides} at higher temperatures, or like doped manganites.\cite{Manganites}

However, a deeper thought of what is known about correlated systems in equilibrium
already raises questions on this point of view. Indeed, according to this picture
one must conclude that energy-pumping, assumed to be equivalent to temperature raising, 
should make a metal less metallic and 
a band insulator less insulating. It is believed\cite{Review_DMFT_96} that a correlated metal near a Mott transition actually shares properties of both metals and insulators; itinerant quasiparticles narrowly centered around the chemical potential coexisting with incoherent atomic-like high-energy excitations, the so-called Hubbard bands. 
If intense light exposure is the same as heating, and since the light 
is selective via its frequency and polarization, then one could envisage  
the quasiparticles or else the Hubbard bands being heated first, which 
would correspond, respectively, to conductivity decrease or increase. 
Such a non-monotonous behavior would however contrast the effect of raising temperature at equilibrium, which is supposed  to always lower conductivity.\cite{Review_DMFT_96}   
The above observation thus challenges the picture of pump-probe experiments as effective thermodynamic perturbations. 

A different perspective, pointing toward an intrinsic kinetic nature of these experimental settings,
is offered by the intense research activity around the non equilibrium dynamics of closed isolated many-body systems, which has recently attracted lot of interest in the different 
context of cold atoms trapped in optical lattices.\cite{SilvaRMP} In this respect, it is by now well established that, when driven out of equilibrium by intense sudden perturbations, strongly correlated systems can be trapped into long-lived metastable states that differ qualitatively from their equilibrium counterpart. Example along this line is provided by the single band Hubbard model, likely the simplest model to describe strong correlation physics. Different theoretical approaches\cite{Kehrein_prl08,Werner_prl09,Werner_long10} have shown, for example, that a sudden increase of the Hubbard repulsion drives the system into a long-lived metastable state which, although highly excited, shows intrinsic features of a zero temperature metallic state, rather than incoherent finite temperature effects as one would have guessed by thermodynamic arguments. Seemingly, suddenly switching on a large Hubbard repulsion stabilizes metastable phases rich of energetically unfavorable doubly occupied sites, which are kinetically blocked\cite{Werner_prl09} and unable to decay\cite{Demler_doublons_prl10} or even to coherently propagate.\cite{Carleo} A qualitative picture of the crossovers or genuine dynamical transitions between different metastable states in the Hubbard model driven by sudden quantum quenches has been recently obtained using a time-dependent extension of the Gutzwiller approximation (t-GA).\cite{SchiroFabrizio_prl10,SchiroFabrizio_PRB11} While missing important quantum fluctuations, which are crucial for the long-time dynamics, this approximate scheme has been shown to capture important qualitative features of the intermediate time evolution, which is actually of interest in the description of the pump-probe dynamics. 

In this work, we aim to elaborate further on this out of equilibrium perspective by including additional ingredients that might play an important role in modeling pump-probe experiments on actual correlated materials. Firstly, we add phonons to the half-filled single band Hubbard model and study the transient dynamics induced by a sudden perturbation. It is worth noticing that lattice vibrations play a crucial role in actual experiments by triggering selective perturbation for the electronic subsystem\cite{Forst_phononics_NatPhys11}, and their role in ultrafast pump probe experiments is a subject of current experimental interest.\cite{Marsi_EPL10,Marsi_PRB10,Caviglia_arXiv11} Here, we consider Einstein phonons coupled to the local charge and study the dynamics of the resulting Hubbard-Holstein model using a suitable extension of t-GA. Although extremely simplified, this model represents a first attempt to figure out how highly excited electrons succeed in transferring their excess energy to the lattice. Results reveal, akin to the pure Hubbard model, the existence of a metastable state for high enough excitation, where phonons get strongly displaced in spite the large Coulomb repulsion and in striking contrast to what one would have guessed in equilibrium.

A second important ingredient that we add to the description builds on the observation, recently reported in a number of theoretical investigations, that non-thermal metastable states are extremely sensitive to spatial fluctuations and prone to spontaneous generation of inhomogeneities,\cite{Foster_prl10,Carleo}  
which could play a crucial role in the dynamics, in particular around dynamical transition points.\cite{SchiroFabrizio_PRB11,SciollaBiroli_long} To investigate this issue, we consider the same Hubbard-Holstein model at half-filling but now in a slab geometry that lacks translational symmetry, modeling ultra-fast dynamics in correlated heterostructures.\cite{Caviglia_arXiv11} We assume that, initially, only the surface layer is driven out of equilibrium and study by t-GA how the excess energy is redistributed inside the bulk. Remarkably, if the energy initially stored on the surface exceeds a critical threshold, it 
remains trapped on the uppermost layers. Concomitantly, the phonons distort at the surface, providing a further trapping potential. This result demonstrates not only the importance of inhomogeneities, but 
also suggests that, under specific circumstances, the lattice might 
not provide a dissipative bath to speed up relaxation, but rather play the opposite game to slow down thermalization.
 
The paper is organized as follows. In section \ref{Method} we introduce the model and an out-of-equilibrium version of the Gutzwiller approximation that may cope with the electron-phonon coupling. In section \ref{Sec:Ising} we show that the method is equivalent to the mean-field approximation applied to a model of free electrons coupled to phonons and Ising spins.
In section \ref{Results} we move to discuss the results for two different cases. First, in section 
\ref{whole-bulk}, we study the time-evolution when the whole bulk is suddenly driven out-of-equilibrium. Next, in section \ref{Surface} we consider the situation in which an external pulse only 
excites the surface layer, and study if and how the surface can relax by transferring energy to the bulk.
Finally, section \ref{Conclusions} is devoted to concluding remarks.

\section{The Model and the Gutzwiller Approximation}
\label{Method}

We consider a half-filled Hubbard-Holstein model described by the 
Hamiltonian 
\ba
H &=& -t\sum_{<ij>\sigma}\,\big(c^\dagger_{i\sigma}c^\dagga_{j\sigma}+H.c.\big)
+ \fract{U}{2}\sum_i\,\big(n_i-1\big)^2\nonumber\\
&& + \frac{\omega}{2}\sum_i\,\big(p_i^2+x_i^2\big) 
-g\sum_i\,x_i\,\big(n_i-1\big),\label{Ham}
\ea
where $c^\dagger_{i\sigma}$($c^\dagga_{i\sigma}$) creates(annihilates) an electron with 
spin $\sigma$ at site $i$, $x_i$ is the phonon displacement at that site and 
$p_i$ its conjugate variable. The hopping is restricted to nearest neighbors 
and $n_i$ is the electron number operator.  
We note that \eqn{Ham} is invariant under particle-hole transformation 
provided $x_i\to -x_i$.

In the following, we study the unitary dynamics induced by the Hamiltonian \eqn{Ham} using the Gutzwiller variational scheme introduced at equilibrium by Barone {\sl et al.}\cite{Barone-PRB} and  
extended to the time dependent case following Ref.~\onlinecite{SchiroFabrizio_prl10}.  
We emphasize at this point that, in real experiments, the light pulse couples via the vector potential 
to the electronic degrees of freedom. While in principle the variational description could be extended to include this feature, here we assume for the sake of simplicity that the effect of the pump is mainly to induce an initial non equilibrium distribution of electronic degrees of freedom, whose dynamics is then driven by the Hubbard-Holstein Hamiltonian.  
Furthermore, while in real experimental settings the system is always in contact with a thermostat that eventually allows the injected energy to flow away, here we assume the whole system made by electrons and lattice to be isolated. While in different contexts, e.g. when current-carrying stationary states driven by static electric fields are present, this assumption may be highly questionable, here we stress that our focus concerns the transient relaxation dynamics on time scales of electronic and phononic degrees of freedom. As the coupling with the environment is typically very weak, we do not believe that this assumption can qualitatively change the physical picture that we will draw.

With these assumptions on the theoretical side, we now introduce our time-dependent variational wave function for the dynamics of the Hubbard-Holstein model.\eqn{Ham} Specifically we write 
\be
\mid \Psi(t)\rangle = 
\prod_i\,\mathcal{P}_i\left(x_i,t\right)\,\mid \Psi_0(t)\rangle,
\label{w-f}
\ee
where $\mid \Psi_0(t)\rangle$ is a time-dependent Slater determinant, to 
be determined variationally, and $\mathcal{P}_i\left(x_i,t\right)$ 
a time-dependent electron operator at site $i$ that depends explicitly 
on the phonon coordinate $x_i$. We define, neglecting the index $i$ for 
convenience,  
\bea
\mathcal{P} &=& \sqrt{2}\;\phi_0(x,t)\,\mid 0\rangle\langle 0\mid
+ \sqrt{2}\;\phi_1(x,t)\,\Big(\mid \up\rangle\langle \up\mid 
+ \mid\down\rangle\langle \down\mid\Big) \nonumber \\
&& + 
\sqrt{2}\;\phi_2(x,t)\,\mid 2\rangle\langle 2\mid,
\label{P}
\eea
where $\phi_n(x,t)$ are site-dependent phonon wave-functions,
and $\mid \Gamma \rangle\langle \Gamma \mid$ is the projector onto 
the site being empty, $\Gamma=0$, singly-occupied by a spin up, 
$\Gamma=\up$, or down, $\Gamma=\down$, electron, or, finally, doubly-occupied, 
$\Gamma=2$. Particle-hole symmetry implies that, under $n\to 2-n$, 
$\phi_n(x,t)\to \phi_{2-n}(-x,t)$, namely
\ba
\phi_0(x,t) &=& \phi_2(-x,t),\\
\phi_1(x,t) &=& \phi_1(-x,t).
\ea
We evaluate average values on the wave-function \eqn{w-f} by means 
of the Gutzwiller approximation, following 
Refs.~\onlinecite{Michele_PRB07_dimer} and \onlinecite{Barone-PRB}, which amounts  to impose that 
\[
\int dx\, \left|\phi_0(x,t)\right|^2 + \left|\phi_1(x,t)\right|^2  = 1. 
\]
The above condition implies that the average over the Slater determinant $\mid \Psi_0(t)\rangle$ 
and the phonons of the operator that remains after extracting from 
$\mathcal{P}_i(x_i,t)^\dagger \mathcal{P}_i(x_i,t)$ any two fermionic operators vanishes identically. 
This property allows to evaluate explicitly all average values on the wavefunction \eqn{w-f} 
in the limit of infinite lattice-coordination,\cite{Gebhard,Michele_PRB07_dimer} although it is common to 
use the same results also for lattices with finite coordination numbers, hence the name {\sl Gutzwiller approximation}.  

Within the Gutzwiller approximation, the average value of the Hamiltonian 
\eqn{Ham} on the wavefunction $\mid \Psi(t)\rangle$ can be shown to coincide with 
the average on $\mid \Psi_0(t)\rangle$ of the Hamiltonian 
\bea
H_*(t) &=& -t\sum_{<ij>\sigma}\, R_i(t)\,R_j(t)\,
\big(c^\dagger_{i\sigma}c^\dagga_{j\sigma}+H.c.\big)\nonumber\\ 
&& + \frac{1}{2}\sum_i\,\int dx\,\Big(U+2gx\Big)\,\left|\phi_{0i}(x,t)\right|^2
 \label{H*}\\
&& + \frac{\omega}{2}\,\sum_i\,\sum_{n=0,1}\,
\int dx\, \phi_{ni}(x,t)^*\,h(x)\,\phi_{ni}(x,t)^\dagga,\nonumber
\eea
where $h(x) = \big(-\partial_x^2 + x^2\big)$. The parameters 
\be
R_i(t) = 
\int dx \Big(\phi_{1i}(x,t)^* \phi_{0i}(x,t)^\dagga + c.c.\Big),
\label{R}
\ee
are commonly interpreted as the amplitudes of quasiparticles at sites $i$, hence 
$H_*$ as their effective non-interacting Hamiltonian 
with renormalized hopping $t_{ij}(t) \equiv R_i(t) R_j(t)\,t$. 

The variational principle that we assume is the saddle point 
of the action $\mathcal{S}=\int dt\,\mathcal{L}(t)$, i.e. 
$\delta \mathcal{S}=0$, whose Lagrangian is 
\be
\mathcal{L}(t) = i\langle\Psi(t)\mid \dot{\Psi}(t)\rangle - 
\langle\Psi(t)\mid \,H\,\mid \Psi(t)\rangle,\label{S1}
\ee
which, within the Gutzwiller approximation,\cite{SchiroFabrizio_PRB11} reads simply 
\bea
\mathcal{L}(t) &=& i\sum_i\,\sum_{n=0,1}\,
\int dx\, \phi_{ni}(x,t)^*\,\dot{\phi}_{ni}(x,t)^\dagga \label{S}\\
&& +i\langle \Psi_0(t)\mid \dot{\Psi}_0(t)\rangle 
- \langle \Psi_0(t)\mid H_*(t)\mid \Psi_0(t)\rangle.\nonumber
\eea
We define on each site $i$ a normalized two-component spinor 
\be
\mid \Phi_i\rangle = \Phi_i(x_i) \equiv
\begin{pmatrix}
\phi_{1i}(x_i)\\
\phi_{0i}(x_i)
\end{pmatrix},\label{Phi}
\ee
so that 
\[
\langle \Phi_i\mid(\dots)\mid \Phi_i\rangle 
= \int dx\, \Phi_i(x)^\dagger\,(\dots)\,\Phi_i(x)^\dagga,
\]
and further introduce Pauli matrices $\sigma^a$, $a=x,y,z$, which act on the 
two components of the spinor. With the above notations, the saddle point equations read:
\bea
i\mid \dot{\Phi}_i\rangle &=& \frac{\omega}{2}\,h\left(x\right) \mid\Phi_i\rangle
-t\sum_{j}^{\text{n.n. of $i$}} R_j\,w_{ij}\,\sigma^x\mid \Phi_i\rangle \nonumber\\
&& +\frac{1}{4}\,\big(U+2gx\big)\,\big(1-\sigma^z\big)\mid \Phi_i\rangle,
\label{eq-Phi}\\
i\mid\dot{\Psi}_0\rangle &=& H_*\mid\Psi_0\rangle,\label{eq-Psi0}
\eea
where
\bea
R_i &=& \langle \Phi_i\mid \sigma^x\mid \Phi_i\rangle, \label{R_i-Ising}\\
w_{ij} &=& \sum_\sigma\,\langle \Psi_0\mid 
\Big(c^\dagger_{i\sigma}c^\dagga_{j\sigma}+H.c.\Big)\mid\Psi_0\rangle.\label{w_ij}
\eea
It is worth noticing here that, when the electron-phonon interaction vanishes, the two subsystems decouple and the above dynamics reduces, for the electronic degrees of freedom, to the one studied in Ref.~\onlinecite{SchiroFabrizio_prl10} for the simple Hubbard model. In the general case, one has to integrate the equations of motion 
starting from initial values for the spinor wave-functions and the Slater determinant. In order to integrate the spinor part, we follow the approach outlined in Ref.~\onlinecite{Barone-PRB} and project each component on the basis of eigenfunctions of the harmonic oscillator, the Hermite functions $\varphi_n(x)$, namely we write for $\nu=0,1$
\be
 \phi_{\nu}(x,t)=\sum_{n=0}^{\infty}\,c_n^{\nu}(t)\,\varphi_n(x)
\ee
and obtain time dependent equations for the complex coefficients $c_n^{\nu}(t)$ by plugging this expansion into equation \eqn{eq-Phi}. In practice, we truncate the basis set to a finite number of coefficients 
$n=0,\dots,N_{b}$ and check that convergence is guaranteed by choosing $N_{b}\simeq 60$.  
All calculations that are presented here have been performed on a cubic lattice using $U=12 t$ and the phonon frequency $\omega=t$. An important scale of energy is the value of the critical $U_c$ at the equilibrium Mott transition in the absence of phonons. In the cubic lattice and within the Gutzwiller approximation $U_c = 16t$, whose inverse we 
shall use as the unit of time. As we mentioned,  the Eqs.~\eqn{eq-Phi}-\eqn{w_ij} are strictly valid only in lattices with infinite coordination numbers, therefore our use in a cubic lattice is just an approximation. 

\subsection{The Gutzwiller approximation as a mean-field theory}
\label{Sec:Ising}
The Eqs.~\eqn{eq-Psi0} and \eqn{eq-Phi} resemble time-dependent mean-field equations, with the Schr{\oe}dinger-like evolution of $\mid \Psi_0\rangle$ that depends implicitly on the average values of selected operators over the wavefunctions $\mid \Phi_i\rangle$, and vice versa for the latter ones.   
Indeed, one recognizes readily that, given the Hamiltonian 
\bea
H_I &=& -t\sum_{<ij>\sigma}\,\sigma^x_i\sigma^x_j\Big(
c^\dagger_{i\sigma}c^\dagga_{j\sigma}+H.c.\Big) \nonumber\\
&& +\fract{1}{4}\,\sum_i\,\Big(U+2gx_i\Big)\,\Big(1-\sigma^z_i\Big)\nonumber\\
&& + \frac{\omega}{2}\,\sum_i\,\Big(p_i^2+x_i^2\Big),\label{HIsing}
\eea
where $\sigma^a_i$, $a=x,y,x$, are Ising variables defined on each site $i$, and assuming a factorized wave-function 
\be
\mid\Psi_I\rangle = 
\mid \text{Ising+phonons}\rangle \times \mid\text{electrons}\rangle, 
\label{Psi-Ising}
\ee
where 
\[
\mid \text{Ising+phonons}\rangle = \prod_i\, \mid \text{Ising+phonons}\rangle_i,
\]
the same variational principle $\delta \mathcal{S}=0$ that we applied before would lead right to Eqs.~\eqn{eq-Psi0} and \eqn{eq-Phi}. The Hamiltonian 
\eqn{HIsing} thus extends to the Hubbard-Holstein model 
the mapping derived in Ref.~\onlinecite{SchiroFabrizio_PRB11} 
for the simple Hubbard model.
We just recall that the mapping states that, 
if $Z$ is the partition function of the original 
model with the Hamiltonian $H$ of Eq.~\eqn{Ham},  
and $Z_I$ that one of the Hamiltonian $H_I$ of Eq.~\eqn{HIsing}, then, 
in the limit of infinite coordination lattices and at particle-hole symmetry,
\be
Z = \left(\frac{1}{2}\right)^N\,Z_I,\label{mapping}
\ee
where $N$ is the number of sites.\cite{SchiroFabrizio_PRB11} Essentially, the mapping demonstrates that the constraint required to implement the so-called slave-spin representation of the Hubbard 
model\cite{DeMedici-1,Z2-1,Z2-2} is actually unessential in the limit of infinite lattice-coordination 
and at particle-hole symmetry. 
The advantage of dealing with 
$H_I$ instead of the original Hamiltonian is that it provides a simple 
framework to disentangle already at the mean field level 
the quasiparticle degrees of freedom, the fermionic operators, 
from the Hubbard bands, the Ising variables. 

The chosen factorization \eqn{Psi-Ising}, where the phonon degrees of freedom 
are entangled with the Hubbard bands and both influence in a mean-field fashion 
the quasiparticles, is actually inspired by the DMFT result that, for 
large repulsion and weak electron-phonon coupling, 
phonon signatures are hardly visible in the 
quasiparticle spectrum but quite evident 
in the Hubbard bands.\cite{Massimo-phonons} Different 
choices could be more appropriate in different contexts or easier to deal with, as the 
extreme factorization $\mid\Psi_I\rangle = \mid \text{Ising}\rangle\times 
\mid \text{phonons}\rangle\times \mid \text{electrons}\rangle$.

\section{Results}
\label{Results}

We shall now analyze the time-dependent mean field equations \eqn{eq-Psi0} 
and \eqn{eq-Phi} that describe within t-GA the evolution of a 
variational wave-function under the action of the Hubbard-Holstein Hamiltonian, 
or equivalently the Hamiltonian \eqn{HIsing}.

We will assume that the pump that drives the system out-of-equilibrium  is selective in the sense 
that it only injects energy in the Hubbard bands, i.e. in the Ising subsystem, 
specifically increasing the concentration of doubly-occupied sites (doublons), 
hence of empty sites (holons) because of particle conservation. 
In the Ising language, it corresponds to assuming that initially the average values of $\sigma^z_i$ are lower than those at equilibrium. We note that the equilibrium conditions are obtained by replacing the time-dependent 
mean-field equations \eqn{eq-Psi0} and \eqn{eq-Phi} with stationary mean-field equations. In particular, the equilibrium values of $\mid\Phi_i\rangle$ are the lowest energy eigenstates of the right hand side of Eq.~\eqn{eq-Phi}, which must be self-consistently determined since the effective Hamiltonian depends on $R_j=\langle \Phi_j\mid\sigma^x\mid\Phi_j\rangle$.\cite{Nicola}

As we mentioned earlier, in real experiments the light pulse couples via the vector potential to both Hubbard bands and quasiparticles. Therefore the above assumption is only an approximation, whose validity we intend to weight up in the near future, while, in the present work, we shall keep assuming that the initial state is just characterized by 
an out-of-equilibrium equal population of doublons and holons. 
We will consider first the case in which such a population is uniformly 
distributed over the whole sample, and next move to inhomogeneous situations.

\subsection{Whole bulk driven out-of-equilibrium}
\label{whole-bulk}

Let us therefore consider the Hubbard-Holstein Hamiltonian \eqn{Ham} at half-filling and assume that 
the system is initially prepared with a uniform concentration of 
doublons and holons higher than at equilibrium. The system is then let evolve, its time evolution being approximated 
within t-GA by Eqs.~\eqn{eq-Psi0} and \eqn{eq-Phi}. This case is actually 
similar to the quench described in Ref.~\onlinecite{SchiroFabrizio_prl10}; 
the new ingredient being just the electron-phonon coupling.

We first need to determine the equilibrium condition in the presence of the electron-phonon coupling. This is accomplished by a self-consistent iterative mapping method similar to the one described in 
 Ref .~\onlinecite{Borghi09}. The outcome is a homogeneous wave-function, with the same spinor at any site, and the Slater determinant that is just the uniform ground state of the hopping energy. 
The concentrations of doublons and holons are then artificially augmented by the same amount while keeping the phonon wave-function unaltered. This is accomplished by the scale transformation 
$\phi_0(x) \to \lambda_0 \phi_0(x)$ and $\phi_1(x)\to \lambda_1 \phi_1(x)$, with $\lambda_0>1$ 
and $\lambda_1<1$ such that  normalization is maintained, 
\[
\int dx\, \lambda_0^2\mid\phi_0(x)\mid^2 + \lambda_1^2\mid\phi_1(x)\mid^2 = 1,
\]
but the concentration of doubly occupied and empty sites is increased. The system is then allowed to evolve as previously explained.  The novelty with respect to Ref.~\onlinecite{SchiroFabrizio_prl10} is that we can now monitor how the  energy, initially injected in the electron subsystem only, is transferred to phonons.
We note that, because translational symmetry is preserved by the time evolution, the effective Hamiltonian $H_*(t)$ in Eq.~\eqn{H*} has $R_i(t)=R_j(t)=R(t)$, $\forall~i,j$, 
 hence describes at all times a simple tight-binding model with uniform time-dependent nearest neighbor hopping. As a result, the Slater determinant that is initially the lowest energy eigenstate of the 
 hopping, does not change in time, hence cannot provide dissipative channels for the spinor evolution. 
 For this reason, the dynamics of both electronic and phononic observables lack relaxation to a steady state but rather shows undamped coherent oscillations. Still, as shown in Ref.~\onlinecite{SchiroFabrizio_prl10}, the mean field dynamics captures important features of the non-equilibrium problem and provides a qualitatively correct picture of the short-to-intermediate time dynamics. 

\begin{figure}[t]
\vspace{-0.8cm}
\centerline{\includegraphics[width=6.3cm,angle=-90]{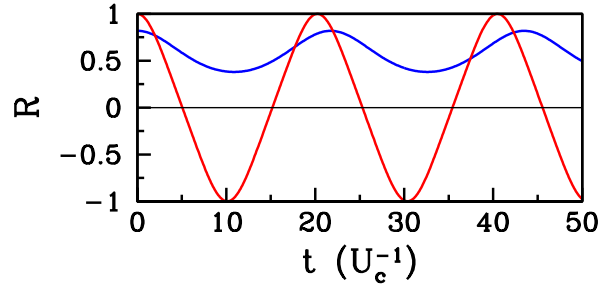}}
\vspace{-1cm}
\caption{The time evolution  
of the parameter $R_i=R$, Eq.~\eqn{R_i-Ising} for two different concentrations $\delta D$ 
of injected doublons, 
one below and the other above the critical point, see Fig.~\ref{fig:D}. Below the critical point, 
$\delta D = 4\%$ (blue curve), $R$ oscillates around a finite value, while, above, $\delta D = 17\%$ (red curve), 
it oscillates between +1 and -1 with zero average.} 
\label{fig:R}
\end{figure}
In Fig.~\ref{fig:R} we plot the time evolution of the renormalization factor $R_i(t)=R(t)$, Eq.~\eqn{R_i-Ising}, which shows two distinct regimes of oscillations depending on the amount of doublons injected, $\delta D$, which measures the strength of the non equilibrium perturbation and that we define as $\delta D\equiv D(t=0)-D_{\text{eq.}}$, with $D(t=0)$ the initial value and $D_{\text{eq.}}$ the equilibrium one.
 
For small perturbations, $R$ oscillates around a finite average, while, upon increasing $\delta D$ above a threshold, it oscillates from -1 to +1, with average zero. In the Ising model language of 
section \ref{Sec:Ising}, this behavior is representative of the transition from the ordered phase, $\langle \sigma^x\rangle\not =0$, to the disordered one, $\langle \sigma^x\rangle=0$. 
A similar dynamical transition was observed in Ref.~\onlinecite{SchiroFabrizio_prl10} for the pure Hubbard model without electron-phonon coupling. The mean field coherent oscillations, although artificial, reflect the real tendency of the system to be trapped into long lived pre-thermal metastable states, whose properties are correctly captured by the long-time averages of the mean field dynamics. 
\begin{figure}[t]
\includegraphics[width=6cm,angle=-90]{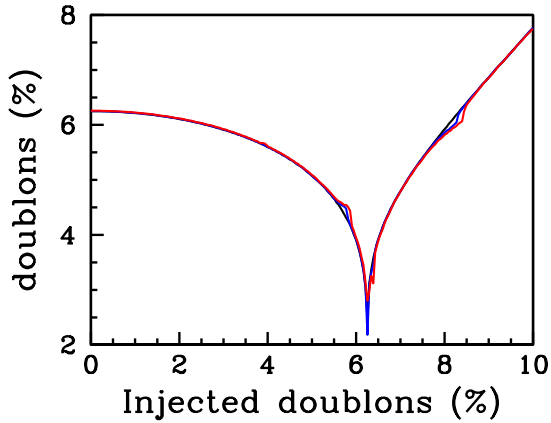}
\caption{The time average value of the percentage of doubly occupied sites 
as function of the percentage $\delta D$ of doublons, hence holons, injected in 
the initial state with respect to the 
equilibrium value.  We plot three values of the electron-phonon couplings, $g=0$ (black), 
$g=0.1t$ (blue) and $g=0.2t$ (red). Larger $g$ correspond to more pronounced kinks near the critical point, which is identified by the point at which the double occupancy drops down.}
\label{fig:D}
\end{figure}

With this insight, we plot in Fig.~\ref{fig:D} the time-averaged double occupancy as a function of the concentration of injected doublons $\delta D$. The first observation is that the electron-phonon interaction does not change qualitatively the behavior with respect to the Hubbard model alone, see Ref.~\onlinecite{SchiroFabrizio_prl10}; namely, we still find two distinct regimes separated by a critical point where the double occupancy goes to zero, although numerically we cannot hit the precise value when this occurs. We note that the location of the critical point is not appreciably affected by phonons because of the tiny electron-phonon coupling ($g^2/\omega \sim 10^{-3}\,U$).  

We find that  the transition occurs right when the initial energy happens to coincide 
with the equilibrium energy at the Mott transition,\cite{SchiroFabrizio_prl10,SciollaBiroli_long}  which is simply the zero point energy of the phonons for our model Hamiltonian \eqn{Ham} and within the Gutzwiller approximation.  In fact, one readily realizes that 
Eqs.~\eqn{eq-Phi} and \eqn{eq-Psi0} admit another stationary point besides the one that corresponds to the equilibrium condition, namely $\phi_0(x)=0$ hence $R=0$, with energy just $\omega/2$.  
We finally mention that, unlike in the absence of phonons,\cite{SchiroFabrizio_prl10} here the critical 
point is not associated to an exponential relaxation towards a 
stationary state that seems to be a characteristic of integrable dynamics,\cite{SciollaBiroli_long} which is presumably not our case. 
 
\begin{figure}[t]
\includegraphics[width=6.3cm,angle=-90]{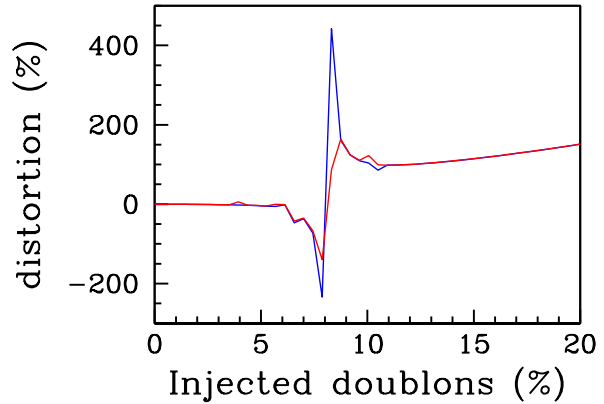}
\caption{The time average value of the lattice distortion, defined as the relative variation with respect to 
the equilibrium value, as function of the percentage of doublons $\delta D$ injected in the initial state with respect to the 
equilibrium value. The curve that is more singular near the critical point corresponds to $g=0.1t$ (blue), 
the other to $g=0.2t$ (red). }
\label{fig:X}
\end{figure}

We now move our attention to the phonon sector, in order to unveil the entanglement between the electrons and the lattice as the former are driven out of equilibrium. A natural quantity to look at would be the average lattice displacement $\langle x_i\rangle$, which however is constrained to be zero on average by particle-hole symmetry.  Still, we can define as a measure of the effective displacement the average of the operator $q_i\equiv x_i (n_i-1)$, which is just the electron-phonon coupling operator. In Fig. \ref{fig:X} we plot the relative variation of the time-averaged effective displacement 
\be
q_* = \lim_{\tau\to\infty} \fract{1}{\tau}\int_0^\tau dt\, \langle  q_i(t)\rangle, 
\label{q_*}
\ee
i.e. $(q_*-q_{\text{eq.}})/q_{\text{eq.}}$, where $q_{\text{eq.}}$ is the equilibrium value, as function of the concentration of injected doublons. We note that, at small concentrations, the displacement 
is mostly unchanged from its equilibrium value. However, for higher concentrations past the critical point, 
the displacement starts increasing substantially;  a growing distortion being a way to store the initial excess energy. 

\begin{figure}[hb]
\includegraphics[width=8.5cm,angle=0]{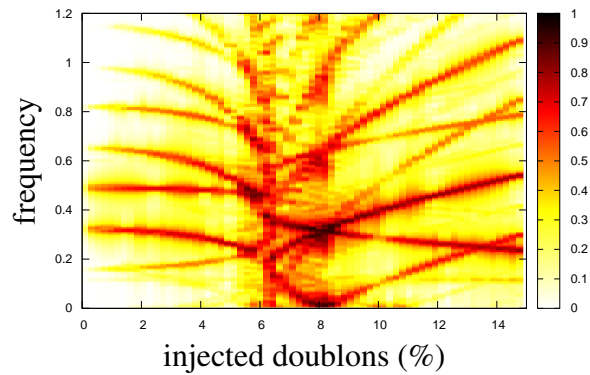}
\caption{Color plot of the Fourier transform (in arbitrary units) of the time evolutions of the double occupancy and the effective phonon distortion, as function of the concentration of injected doublons 
and of the frequency in units of $U_c$.}
\label{fig:frequency}
\end{figure}
Although the gross behavior seems not to be affected by phonons, there are details which feature their  
presence. In particular, we note some anomalies, tinier in Fig.~\ref{fig:D} and more visible in  
\ref{fig:X}. These anomalies appear when the oscillation frequency of the electronic dynamics, which decreases on approaching the critical  point, hits the renormalized phonon frequency, or a multiple of it.  
Since the latter is small, these resonances occur near the critical point.  
This is evident in Fig.\ref{fig:frequency}, where 
we draw by a color plot the spectral decomposition of the time evolutions both of the double occupancy and 
the phonon distortion as function of the concentration of injected doublons and of the frequency of the signal. In particular, we observe the avoided crossing between the two lowest frequencies 
around $\delta D\sim 5\%$ that causes the kink visible in Fig.\ref{fig:D}. Among these two frequencies, the lowest one is visible mostly in the dynamics of $q(t)$, hence can be regarded as a {\sl renormalized} phonon frequency that blue-shifts upon increasing $\delta D$, i.e. the energy injected into the system. 

\subsection{Surface driven out-of-equilibrium}
\label{Surface}

\begin{figure}
\centerline{\includegraphics[width=6cm,height=4cm]{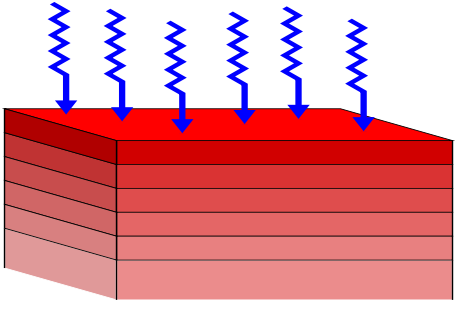}}
\caption{Slab geometry for simulating the hypothetical experiment in which only the surface layer 
is driven out of equilibrium}
\label{geometry}
\end{figure}
Let us now consider a slab geometry as depicted in Fig.~\ref{geometry}, denoting by $z$ the direction 
perpendicular to the surface, which lies in the $xy$-plane. This setting allows us to mimic the non-equilibrium dynamics across a correlated heterostructure, that has recently attracted experimental interest.\cite{Caviglia_arXiv11} We consider a system described by the Hubbard-Holstein model and consider a perturbation acting only at the surface layer, by triggering an out-of-equilibrium population of doublons and holons, while keeping the bulk in its equilibrium ground-state configuration. This initial state is then let evolve and its time-evolution is approximated by the Eqs.~\eqn{eq-Psi0} and \eqn{eq-Phi}. 

This particular geometry has the additional complication that, at equilibrium, the optimized $\mid \Phi_i\rangle$ are layer dependent and the optimized Slater determinant is not uniform anymore. 
Therefore, the first step we need to undertake is solving the equilibrium problem, which we accomplish by the method developed in Ref.~\onlinecite{Borghi09}. Because of the slab geometry, we can choose 
a basis of single-particle wave-functions for building the Slater determinant defined by 
\[
\psi_{\epsilon\bk}(\br,i) = \fract{\text{e}^{i\bk\cdot\br}}{\sqrt{A}}\;\psi_{\epsilon\bk}(i),
\]
where $\br$ is the space coordinate and $\bk=(k_x,k_y)$ the momentum in the $xy$-plane, which is assumed to contain $A$ lattice sites, while $i=1,\dots,N$ is the layer index, and typically we used $N=100$ layers. 
Since there is translational symmetry in the $xy$-plane, we can choose the spinor $\mid \Phi_i\rangle$ to depend only on the layer index $i$. 
Then, the stationary solution of \eqn{eq-Phi} and \eqn{eq-Psi0} amounts to solve at fixed $R_i$ the eigenvalue problem 
\be
\epsilon\,\psi_{\epsilon\bk}(i) = t\,R_i^2\,\epsilon_\bk\,\psi_{\epsilon\bk}(i) 
-t\,R_i\,\sum_{a=\pm1} \,R_{i+a}\,\psi_{\epsilon\bk}(i+a),\label{step-1}
\ee
where $\epsilon_\bk = -2\left(\cos k_x + \cos k_y \right)$, with  the boundary condition 
$\psi_{\epsilon\bk}(0)=\psi_{\epsilon\bk}(N+1)=0$. The lowest energy eigenfunctions 
$\epsilon<\epsilon_F$, $\epsilon_F=0$ because of particle-hole symmetry,  are then used to define the Slater determinant $\mid \Psi_0\rangle$ and the average hopping between layer $i$ and $i+a$
\be
w_{i\to i+a} = \frac{1}{A}\sum_{\epsilon<0}\sum_{\bk  \ni \epsilon<0}\,\Big(
\psi_{\epsilon\bk}(i)^*\psi_{\epsilon\bk}(i+a)^\dagga + c.c.\Big),\label{w-i-to-i+a}
\ee
as well as the average hopping within layer $i$
\be
w_{i\to i} = \frac{1}{A}\sum_{\epsilon<0}\sum_{\bk  \ni \epsilon<0}\,\epsilon_\bk\,
\mid\psi_{\epsilon\bk}(i)\mid^2.\label{w-i-to-i}
\ee
\begin{figure}[t]
\centerline{\includegraphics[width=6cm,angle=-90]{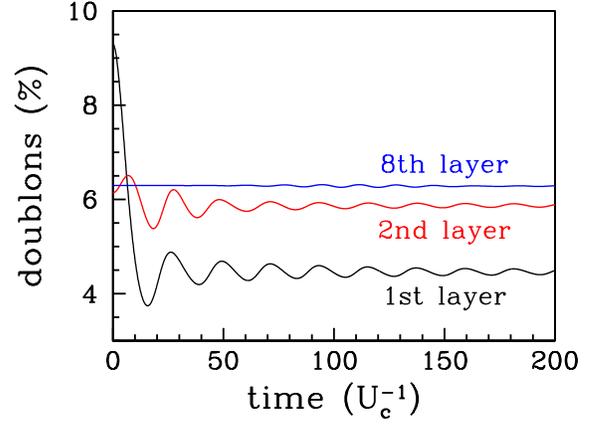}}
\caption{Time evolution of the percentage of doubly occupied sites on three different layers,  indicated in the figure, for small percentage of injected doublons. The time is measured in units of the inverse of $U_c$. The simulation is performed with a slab of 100 layers.}
\label{D_below}
\end{figure}
These parameters are used to solve the spinor eigenvalue problem 
\bea
E\mid\Phi_i\rangle &=& \frac{\omega}{2}\,h(x) \mid\Phi_i\rangle 
-t R_i\,w_{i\to i}\,\sigma^x\mid\Phi_i\rangle \nonumber\\
&& -t \sum_{a=\pm 1}\, R_{i+a}\,w_{i\to i+a}\,\sigma^x\mid\Phi_{i+a}\rangle \nonumber\\
&& + \frac{1}{4}\left(U+2gx\right)\left(1-\sigma^z\right)\mid\Phi_i\rangle,\label{step-2}
\eea
whose lowest energy solution defines new parameters $R_i=\langle \Phi_i\mid 
\sigma^x\mid\Phi_i\rangle$ that are used to solve again Eq.~\eqn{step-1} and so on, until 
convergence is reached.\cite{Borghi09,Nicola} 

The breaking of translational symmetry by the presence of the surfaces is actually amplified by 
electron correlations that create a surface {\sl dead layer},\cite{Marsi,Borghi09} with suppressed double occupancy, hence reduced hopping renormalization parameters $R_i$. The dead layer penetrates inside the bulk over a length proportional to the Mott-transition correlation length.\cite{Borghi09,Marsi} 

Given this starting state, we suddenly increase the population 
of doublons and holons on the first layer $i=1$ and let the system evolve. Essentially, we simply turn Eqs.~\eqn{step-1} and \eqn{step-2} into self-consistent non-linear time-dependent Schr{\oe}dinger equations that we solve numerically. Unlike in the homogeneous case of section 
\ref{whole-bulk},  here the Slater determinant evolves with time because the wave-functions 
$\psi_{\epsilon\bk}(i,t)$ acquire a non-trivial time dependence, which provides additional dissipative channels that were previously absent. In other words, the hopping parameters $w_{i\to i+a}(t)$ and $w_{i\to i}(t)$ 
defined in Eqs.~\eqn{w-i-to-i+a} and \eqn{w-i-to-i} become time-dependent and influence the evolution 
of $\mid \Phi_i\rangle$, see Eq.~\eqn{step-2}, which in turns  affects $\psi_{\epsilon\bk}(i,t)$ via 
the parameters $R_i(t)$. 

 \begin{figure}[t]
\centerline{\includegraphics[width=6cm,angle=-90]{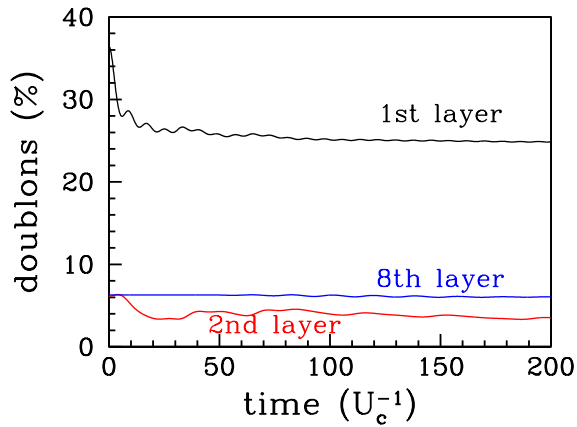}}
\caption{Same as Fig.~\ref{D_below} at higher percentage of injected doublons.}
\label{D_above}
\end{figure}
The mutual feedback between $\mid\Psi_0\rangle$ and 
the $\mid\Phi_i\rangle$'s brings about a non-trivial dynamics much richer than in the example 
discussed in section \ref{whole-bulk}. Nevertheless, even in this case we do find two completely different dynamical behaviors, depending on the amount of injected doublons. For small values, the perturbed surface layer is able to relax 
by dissipating its excess energy inside the bulk, see Fig.~\ref{D_below}. 
Indeed, the time-average values of the double occupancies on each layer tend towards their equilibrium 
values, which, as we mentioned, are lower the closer the layer to the surface. We emphasize that here, as opposite to the previous case and to the case of global quantum quenches, the perturbation is local and the energy injected does not scale with the system size. As a result, the relaxation dynamics we find in this regime is toward the equilibrium ground state and no heating or finite temperature effects are expected in the long-time limit. This is a specific example of a local quantum quench and shows that our time dependent Gutzwiller approximation, with the above mentioned feedback between variational parameters and Slater determinant, is able to describe thermalization. 
We also mention that, working with a finite-size geometry, recurrence effects are present at long enough times, when the perturbation reaches the opposite surface and starts oscillating back and forth. We expect that by taking the thermodynamic limit in the $z$ direction these finite-size effects will be washed away. Nevertheless, even for finite lengths the relaxation and the trend towards equilibrium are clearly evident.

\begin{figure}[t]
\centerline{\includegraphics[width=6cm,angle=-90]{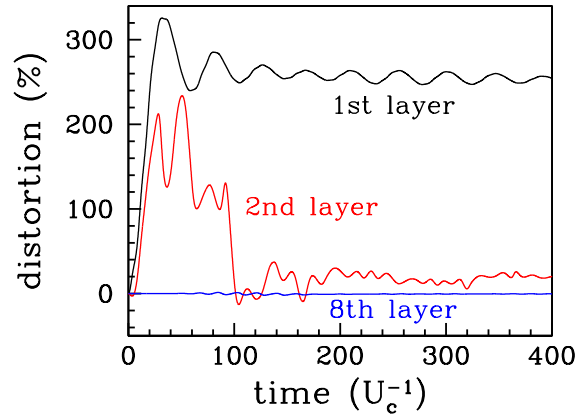}}
\caption{The lattice distortion on three different layers with the same amount of injected doublons as in 
Fig.~\ref{D_above}.}
\label{X_above}
\end{figure}
Upon increasing the concentration of injected doublons a different dynamical behavior emerges. In particular, above a certain threshold the excitation remains trapped near the surface, see Fig.~\ref{D_above}. This is quite remarkable because, as we mentioned, the Slater determinant now adjusts to the spinors $\mid \Phi_i\rangle$ during the time evolution, hence could in principle absorb the 
excess energy and transfer it in the interior of the bulk. What actually happens 
is that the parameters $R_i$ and $R_{i+1}$ of adjacent layers interfere destructively, i.e. oscillate out-of-phase, at some $i$ near the surface, effectively suppressing the quasiparticle inter-layer hopping 
$t_{ii+1}=R_i\,R_{i+1}\,t$, hence cutting layer $i$ from the rest of  the bulk. 
This anomalous trapping exists also in the absence of electron-phonon coupling, hence it is primarily an electronic effect, presumably the dynamical counterpart of the surface {\sl dead layer} at equilibrium.\cite{Marsi,Borghi09} What changes 
at finite electron-phonon coupling is that this phenomenon is accompanied by a lattice deformation, also localized on the uppermost layers, see Fig.~\ref{X_above}. 

\begin{figure}[t]
\vspace{-0.4cm}
\centerline{\includegraphics[width=6cm,angle=-90]{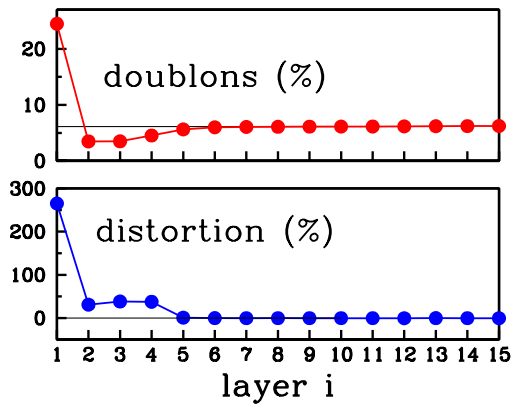}}
\caption{The percentage of doubly occupied sites and of the lattice distortion at each layer at very long times.}
\label{xd}
\end{figure}

\begin{figure}[t]
\centerline{\includegraphics[width=6cm,angle=-90]{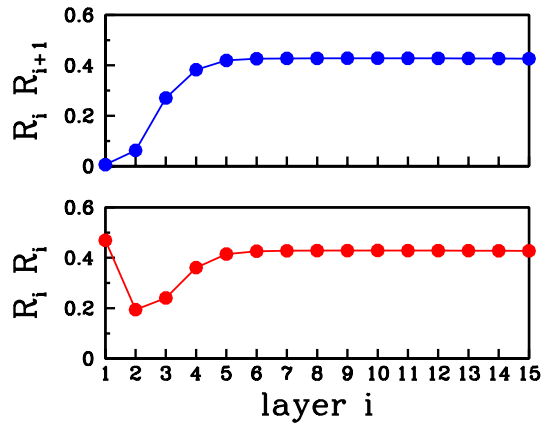}}
\caption{The intra-layer, bottom panel, and inter-layer, top panel, hopping renormalization factors at long times.}
\label{Zsurf}
\end{figure}

The physical picture that emerges can be visualized much better by the long-time layer-dependent profiles of the percentages of doubly occupied sites and of the distortion, shown in Fig.~\ref{xd}. 
We observe that the deviations of both quantities with respect to equilibrium are indeed concentrated just near the surface, while the bulk is practically unaffected. Also instructive 
are the profiles of the intra-layer and inter-layer hopping renormalization factors, $R_i^2$ and $R_i\,R_{i+1}$, shown in 
Fig.~\ref{Zsurf}. We note that the first layer has a much larger hopping 
renormalization factor,  $R_1^2$, than at equilibrium, when it would be very small due to the {\sl dead layer} phenomenon.\cite{Marsi,Borghi09}  However, this layer is practically decoupled from the second layer, $R_1\,R_2$ being vanishingly small. 

We end by mentioning that, in contrast to the case of section \ref{whole-bulk},  
the two different regimes that we observe seem not to be separated by a genuine dynamical critical point, but rather by the dynamical counterpart of a first order phase transition. 
Indeed, in the intermediate regime the system does not show a well defined behavior but instead oscillates between the two distinct phases above. 

\section{Conclusions}
\label{Conclusions}

In this work we have studied the real time dynamics of the Hubbard-Holstein model at half-filling 
by a very simple extension of the Gutzwiller approximation in two different toy cases: 
({\sl i}) a bulk system is prepared with an equal out-of-equilibrium population of doubly occupied 
and empty sites and let evolve in time; ({\sl ii}) a slab is considered and it is assumed that only the surface layer is initially driven out-of-equilibrium. 

In case ({\sl i}) we find similar results as in the quantum quench of the pure Hubbard model: a dynamical critical point that separates two different regimes. The novel feature introduced by the phonons is the presence of a substantial phonon-displacement that occurs for large enough deviation from equilibrium, a remarkable outcome in that the 
electron-phonon coupling we consider is extremely small as compared with the Hubbard repulsion. 

In the slab geometry ({\sl ii}) we still find two dynamical behaviors, although this time without a true dynamical transition in between. If the energy injected at the surface is below a threshold, 
it is able to diffuse within the bulk and the system seems to relax towards the equilibrium ground state  with the {\sl dead layer} near the surface.\cite{Marsi,Borghi09} 
On the contrary, if the excess energy at the surface exceeds that threshold, it does not succeed anymore 
to diffuse in the bulk and remains concentrated practically at the surface, 
bringing about a substantial phonon displacement. Surprisingly, we finds that the first layer has a larger hopping renormalization factor than at equilibrium, which can be sustained because the layer 
effectively decouples from the rest of the system. However, we can not conclude that such an enhancement corresponds to an increased metallicity, which would be indeed a remarkable result. In fact, we tend to believe that the hopping renormalization as defined within the Gutzwiller approximation is 
a measure of the whole, coherent plus incoherent, single-particle spectrum at low energy, not just of the 
quasiparticle coherent contribution alone. Therefore, what we feel safe to state is just that low energy spectral weight grows in the first layer, whatever being its nature. 
    
We cannot exclude that such  a long-lived localized excitation could indeed correspond to some kind of exciton already present in the equilibrium spectrum, which can be unveiled by our variational technique only because we are exploring the dynamics. 
It is also plausible that such a localized excitation exists just in correspondence with the 
surface {\sl dead layer},\cite{Marsi, Borghi09} where the low-energy spectral weight is 
negligible hence there is room for excitons inside the preformed Mott-Hubbard gap. 
It is as well possible that our finding is actually related to the debated issue about the 
lifetime of doublons in the strongly interacting Hubbard model,\cite{Demler_doublons_prl10,Shastry,Eckstein_prb11} 
which could also be the clue to understand the lack of thermalization of highly excited states when 
correlation is strong. Further investigations with different and complementary approaches are needed to clarify these interesting issues.
 
 \section*{Acknowledgments}
 This work has been partly supported by PRIN/COFIN 20087NX9Y7 and by EU/FP7 under the project LEMSUPER.  
    

\end{document}